%% file: main.tex
\documentclass[10pt,conference]{IEEEtran}
\usepackage[T1]{fontenc}

\AtBeginDocument{%
  \providecommand\BibTeX{{%
    Bib\TeX}}}

\makeatletter
\newcommand{\linebreakand}{%
  \end{@IEEEauthorhalign}
  \hfill\mbox{}\par
  \mbox{}\hfill\begin{@IEEEauthorhalign}
}
\makeatother

\input{macros}

\begin{document}


\title{\approach: Context-aware, Repository-level Code Translation with Isolated Validation}

 \author{
   \IEEEauthorblockN{Kaiyao Ke}
   \IEEEauthorblockA{UC Berkeley\\kaiyaoke@berkeley.edu}
   \and
   \IEEEauthorblockN{Ali Reza Ibrahimzada}
   \IEEEauthorblockA{University of Illinois Urbana-Champaign\\alirezai@illinois.edu}
   \and
   \IEEEauthorblockN{Rangeet Pan}
   \IEEEauthorblockA{IBM\\rangeet.pan@ibm.com}
   \and
   \IEEEauthorblockN{Saurabh Sinha}
   \IEEEauthorblockA{IBM\\sinhas@us.ibm.com}
   \linebreakand
    \IEEEauthorblockN{Reyhaneh Jabbarvand}
   \IEEEauthorblockA{University of Illinois Urbana-Champaign\\reyhaneh@illinois.edu}
 }


\maketitle

\begin{abstract}
  \input{Resources/Sections/0-Abstract}
\end{abstract}

\input{Resources/Sections/1-Introduction}
\input{Resources/Sections/2-Illustrative-Example}
\input{Resources/Sections/3-Overview}
\input{Resources/Sections/4-Approach}
\input{Resources/Sections/5-Evaluation}

\input{Resources/Sections/6-Related-Work}
\input{Resources/Sections/7-Threats-to-Validity}
\input{Resources/Sections/8-Conclusion}
\input{Resources/Sections/9-Artifacts}

\section*{Acknowledgment}
This work is supported by the NSF CCF-2238045 grant. We thank Prof. Darko Marinov for his continuous support throughout this project and for guidance on handling edge cases in the Java serializer.

\bibliographystyle{IEEEtran}
\bibliography{main}

\end{document}

%% file: macros.tex
\usepackage[numbers,sort&compress]{natbib}
\usepackage{minted}
\usepackage[noend]{algpseudocode}
\usepackage{graphicx}
\usepackage[tikz]{bclogo}
\usepackage{wrapfig}
\usepackage{textcomp}
\usepackage{xcolor}
\usepackage{arydshln}
\usepackage{enumitem}
\usepackage{soul}
\usepackage[belowskip=0pt,aboveskip=5pt]{subcaption}
\usepackage{listings}
\usepackage{xspace}
\usepackage[most]{tcolorbox}
\usepackage{multirow}
\usepackage{color}
\usepackage{url}
\usepackage[export]{adjustbox}
\usepackage[linesnumbered,ruled,vlined]{algorithm2e}
 \usepackage{booktabs}

\usepackage{amsmath,amssymb,amsfonts}
\def\BibTeX{{\rm B\kern-.05em{\sc i\kern-.025em b}\kern-.08em
    T\kern-.1667em\lower.7ex\hbox{E}\kern-.125emX}}

\usepackage{cleveref}
\definecolor{bubblegum}{rgb}{0.99, 0.76, 0.8}
\definecolor{cambridgeblue}{rgb}{0.64, 0.76, 0.68}
\newcommand{\approach}{\textsc{MockingBird}\xspace}

\newcommand{\alphatrans}{\textsc{AlphaTrans}\xspace}
\definecolor{ibmblue}{RGB}{63,97,246}

\newcommand{\rangeet}[1]{\textcolor{blue}{\textbf{Rangeet:} #1}}

\setmintedinline{breaklines}

\SetKwInput{KwInputs}{Inputs}                
\SetKwInput{KwOutputs}{Outputs}              
\SetKwInput{KwOutput}{Output}              

\def\BibTeX{{\rm B\kern-.05em{\sc i\kern-.025em b}\kern-.08em
    T\kern-.1667em\lower.7ex\hbox{E}\kern-.125emX}}

\newcommand{\mybox}[1]{\begin{tcolorbox}[enhanced, frame hidden, boxsep=0pt]\emph{#1}\end{tcolorbox}}

\definecolor{problemblue}{RGB}{100,134,158}
\definecolor{idiomsgreen}{RGB}{0,162,0}
\definecolor{exercisebgblue}{rgb}{0,  .69,  .941}
\definecolor{deepgreen}{rgb}{0.0, 0.5, 0.0}
\definecolor{codegreen}{rgb}{0,0.6,0}
\definecolor{codegray}{rgb}{0.5,0.5,0.5}
\definecolor{codepurple}{rgb}{0.58,0,0.82}
\definecolor{backcolour}{rgb}{0.95,0.95,0.92}
\definecolor{redColor}{RGB}{255,0,0}
\definecolor{Gray}{gray}{0.1}

\lstdefinestyle{mystyle}{
	backgroundcolor=\color{backcolour},   
	commentstyle=\color{codegreen},
	keywordstyle=\color{magenta},
	numberstyle=\tiny\color{codegray},
	stringstyle=\color{codepurple},
	basicstyle=\scriptsize,
	breakatwhitespace=false,         
	breaklines=true,                 
	captionpos=b,                    
	keepspaces=false,                 
	numbers=left,                    
	numbersep=5pt,                  
	showspaces=false,                
	showstringspaces=false,
	showtabs=false,                  
	tabsize=2, numbers=left,
    breaklines=true,
    rulecolor=\color{black}
}
\lstdefinelanguage{test}{%
	language     = python,
	breaklines = true,backgroundcolor=\color{white},escapechar=!,rulecolor=\color{black}, breaklines=true,sensitive=true,  numbersep=5pt, xleftmargin=.015\textwidth, frame=tb,label=test
}
\lstdefinelanguage{source}{%
	language     = python,
	breaklines = true,
firstnumber=0,numberfirstline=false,columns=fullflexible,numbers=left,backgroundcolor=\color{white},
    rulecolor=\color{black}, 
    morecomment=[f][\color{codegreen}]{+\ },
    morecomment=[f][\color{redColor}]{-\ },    
    breaklines=true,sensitive=true, numbersep=5pt, xleftmargin=.015\textwidth, label=test
}
\newcommand*\Suppressnumber{%
  \lst@AddToHook{OnNewLine}{%
    \let\thelstnumber\relax%
     \advance\c@lstnumber-\@ne\relax%
    }%
}

\lstdefinestyle{customc}{
	belowcaptionskip=1\baselineskip,
	breaklines=false,
	frame= single,
	breaklines = true,
	xleftmargin=\parindent,
	language= Python,
	showstringspaces=false,
	basicstyle=\footnotesize\ttfamily,
	keywordstyle=\bfseries\color{green!40!black},
	commentstyle=\itshape\color{purple!40!black},
	identifierstyle=\color{blue},
	stringstyle=\color{codegreen},
	backgroundcolor=\color{gray!4}
}

\DeclareCaptionFormat{listing}{#3}
\newcommand{\killpunct}[1]{}

%% file: Resources/Sections/0-Abstract.tex
Despite advances in repository-level code translation, automated type resolution across differing type systems and translation validation (functional-equivalence checking) remain challenging. This paper proposes \approach, combining context-aware type resolution with mock-based isolated validation. \approach retrieves API documentation and contextual code information per source-language type, prompting an LLM to resolve precise cross-language type mappings, reducing the manual effort of existing approaches. Using this automatically constructed type map, \approach employs custom serialization/deserialization to generate equivalent target-language mock objects, validating each translation unit \emph{in isolation} without \emph{test translation}, eliminating test-overfitting risks and translation challenges.
On 10 Java-to-Python tasks from the \alphatrans benchmark, \approach improves functional equivalence from the state-of-the-art 25.14\% to 52.8\%, with no manual type-mapping correction needed. A 700-fragment study shows \approach achieves 0.1\% false positives and 0.3\% false negatives, substantially outperforming GraalVM (2.1\% false positives, 14.0\% false negatives) and MatchFixAgent (11.0\% combined mislabel rate). The pipeline is also LLM-agnostic, maintaining these gains with both DeepSeekCoder-33B and the smaller Devstral-24B.

%% file: Resources/Sections/1-Introduction.tex
\section{Introduction}
\label{sec:introduction}

Repository-level code translation is critical for software sustainability and modernization. Existing recent such techniques fall into two categories: neuro-symbolic~\cite{ibrahimzada2025alphatrans,nitin2025c2saferrust,shetty2024syzygy,wang2025evoc2rust,cai2025rustmap,zhang2024scalable,zhou2025llm,wang2025program,luo2025integrating} and agentic~\cite{sim2025large,ibrahimzada2026matchfixagent,yuan2025project,khatry2025crust,wang2025effireasontrans,li2025adversarial,guan2025repotransagent,ibrahimzada2026recodeagent,farrukh2026orbit}. Neuro-symbolic approaches combine LLMs with program analysis in a fixed pipeline; agentic approaches let LLM agents autonomously explore and translate using tools.

A critical shared component is \emph{translation validation} to ensure functional equivalence with the source. Formal methods do not scale to whole repositories~\cite{yang2024vert}. Test-based validation (translating/executing existing tests~\cite{ibrahimzada2025alphatrans,khatry2025crust}, differential fuzzing~\cite{zhang2024scalable,bai2025rustassure}) suffers from: (1) \emph{test coupling effect}~\cite{ibrahimzada2025alphatrans}, which results one bug in dense intra/inter-procedural dependencies invalidates the whole call chain translation; and (2) \emph{test/code overfitting}~\cite{chen2026unlocking}, where LLM-translated tests and code can overfit to each other, passing without true validation.

\alphatrans~\cite{ibrahimzada2025alphatrans} introduced \emph{isolated testing} via GraalVM~\cite{graalvm}, running source-language tests on individual translated target fragments. While this overcomes \emph{test coupling effect} and \emph{test/code overfitting}, it suffers from (1) \emph{incompleteness} from GraalVM constraints; (2) \emph{lack of scalability}, requiring semi-automated glue code generation and type mapping; (3) \emph{incompatibility} with automated test generation (GraalVM's Java version constraints); (4) \emph{false negatives} (e.g., glue-code type mismatches triggering spurious \texttt{\small{NullPointerException}}); and (5) \emph{false positives} (e.g., undetected global iterator side effects).

MatchFixAgent~\cite{ibrahimzada2026matchfixagent} uses a frontier LLM (Claude Sonnet 3.7) to judge and repair equivalence, avoiding GraalVM pitfalls. This approach is costly. Furthermore, using an LLM in the pipeline for test generation, translation, and validation makes it susceptible to the \emph{test/code overfitting} mentioned earlier, leading to false positives. In fact, of 758 fragments validated by both \alphatrans and MatchFixAgent, 233 disagreed, and manual inspection confirmed \alphatrans was correct in 62 of them, lower-bounding MatchFixAgent's error rate at $62/758 = 8.18\%$ (as we will show, on three projects covered by our manual study, MatchFixAgent exhibits an 11.0\% mislabeling). 

We propose \approach for \emph{mock-based isolated validation} to overcome the limitations mentioned in prior work. It executes source-language tests, capturing I/O pairs and side effects (field mutations, global/static state changes, control-flow-altering exceptions) per method invocation, then \emph{serializes} this data. A \emph{deserialization} step generates mocks for nested invocations, reconstructing target-language tests for equivalence checking, \emph{without test translation or cross-language execution}. Serialization/deserialization uses \emph{context-aware} type mapping: \approach retrieves API documentation and code context per type, then uses RAG-based LLM prompting to resolve cross-language type semantics.

\begin{figure*}[t]
    \centering
    \includegraphics[width=0.9\textwidth]{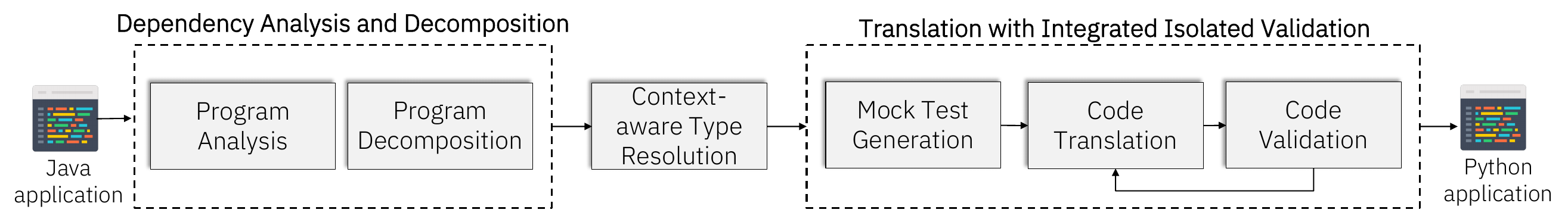}
    \caption{Overview of \approach}
    \vspace{-10pt}
    \label{fig:overview}
\end{figure*}

\approach applies to language pairs where (1) the source has good type/API documentation for RAG-based resolution, and (2) the target has a mature mocking framework. We chose Java$\rightarrow$Python\footnote{Please see \S \ref{sec:threats} for generalizability discussion.} for evaluation due to their notably divergent, complex type systems, making it a harder setting that better demonstrates \approach's power. Real-world demand is exactly what motivated \alphatrans, MatchFixAgent and \approach: many influential Java systems were re-implemented in Python because no reliable automated translation existed---e.g., Python's standard-library \texttt{\small{unittest}} began as PyUnit, a manual port of Java's JUnit; Stanford CoreNLP was re-implemented by the same group as Stanza; Apache Lucene inspired the pure-Python Whoosh. That these were re-implementations rather than automatic translations shows that Java$\rightarrow$Python translation has been too hard to automate; even the state of the art reached only ${\sim}25\%$ validated equivalence, which is why we target this pair rather than better-served pairs like C$\rightarrow$Rust. Furthermore, this translation pair allows us to directly compare with alternative neuro-symbolic and agentic approaches across real-world projects and different techniques for in-isolation validation, namely Graal-based validation by \alphatrans and unit test generation by MatchFixAgent. 

On these ten diverse Java projects (including concurrency-heavy types), \approach achieves $52.8\%$ functional equivalence versus \alphatrans's $25.14\%$, with substantially fewer false positives/negatives than GraalVM-based validation.
Compared to MatchFixAgent~\cite{ibrahimzada2026matchfixagent}, the state-of-the-art LLM-based validation and repair framework, \approach provides higher coverage, higher reliability, and lower cost. MatchFixAgent's reported 75.9\% equivalence is measured on only 4 of the 10 \alphatrans benchmark projects, where \alphatrans+GraalVM already achieves 42.6\% (vs.\ 25.1\% across all 10 projects). Moreover, its reported 8.18\% combined false-positive/false-negative rate is non-negligible. MatchFixAgent also incurs \$1{,}642 in Claude API costs, on average, only to validate and repair 1{,}346 translation fragments in four projects, whereas \approach scales to all 10 projects (4,643 translation fragments) using open-source models (DeepSeekCoder-33B-Instruct and Devstral-24B), achieving only 0.4\% combined mislabels without incurring per-token API costs. \approach's performance also generalizes across LLMs, including smaller models like Devstral-24B. Our contributions are: (1) a novel automated pipeline for mock-based isolated validation requiring no test translation or cross-language execution, producing equivalent target-language tests; and (2) comprehensive evaluation against state-of-the-art neuro-symbolic and agentic approaches, showcasing the superiority of \approach.

%% file: Resources/Sections/3-Overview.tex
\section{Overview}
\label{sec:overview}

Figure~\ref{fig:overview} presents \approach's three main components: \textit{Dependency Analysis and Decomposition} (\S\ref{subsec:transformation-decomposition}), \textit{Context-aware Type Resolution} (\S\ref{subsec:type-resolution}), and \textit{Translation with Integrated Isolated Validation} (\S\ref{subsec:compositional-translation-validation}). The first two decompose the source program and resolve types; the third translates and validates decomposed units in isolation, then recomposes the full target-language program.

\textit{Dependency Analysis and Decomposition} breaks a Java project into field- or method-level translation units (\emph{fragments}), collecting caller/callee dependencies into a JSON \emph{schema} used by later components (\S\ref{subsec:transformation-decomposition}). It also runs existing tests to gather fragment coverage, using EvoSuite to augment unit tests to maximize coverage.
\textit{Context-aware Type Resolution} takes the schema and resolves library types from the source to functionally equivalent target types. It crawls source-type API documentation, gathers surrounding code context, and prompts an LLM with this information to predict the equivalent type, validating predictions via a two-step approach (\S\ref{subsec:type-resolution}).

\textit{Translation with Integrated and Isolated Mock-based Validation} takes the type-resolved schema and translates fragments in reverse topological-dependency order using an LLM. \approach generates mock tests from the source project's existing tests, isolating each focal fragment by mocking other fragments' behavior, and builds the harness to run these mock tests in the target PL. After prompting the LLM to translate each fragment and checking syntactic correctness, it executes mock tests to verify functional equivalence; on failure, it iteratively re-prompts the LLM with the failure trace and task context, up to a bounded number of repair attempts (\S\ref{subsec:compositional-translation-validation}).

%% file: Resources/Sections/4-Approach.tex
\section{\approach Framework}
\label{sec:approach}

\begin{algorithm}[t]
\scriptsize
\input{Resources/Algorithms/type-resolution}
\end{algorithm}

\subsection{Dependency Analysis and Decomposition}
\label{subsec:transformation-decomposition}

\begin{figure*}[t]
    \centering
    \includegraphics[width=0.8\textwidth]{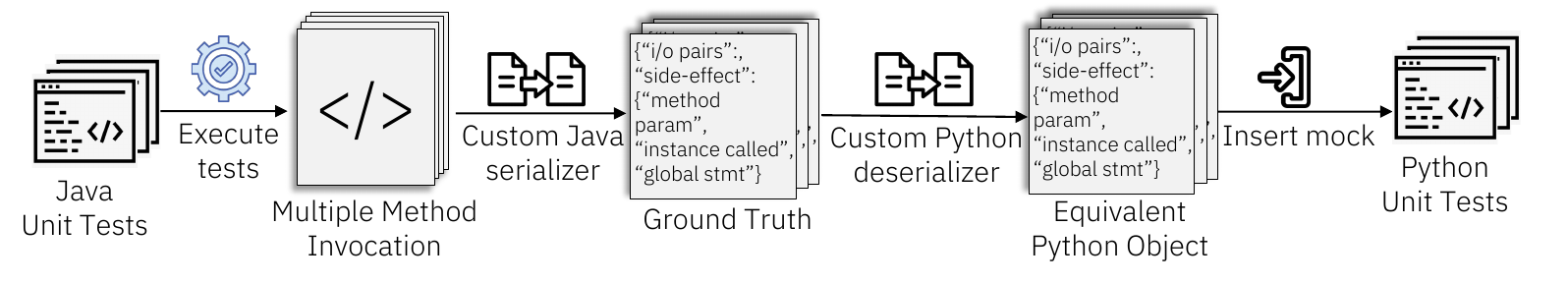}
    \vspace{-10pt}
    \caption{Mock-based Isolated Validation}
    \label{fig:mocking}
    \vspace{-15pt}
\end{figure*}

Real-world projects exceed LLM context windows; even frontier models that nominally fit a repository degrade before reaching the limit, with cost scaling quadratically with sequence length~\cite{liu2023positionalinterpolation,ibm2024contextwindow}. Decomposition instead focuses attention on relevant information, reducing noise and hallucinations while improving accuracy~\cite{wang2025document} and enabling focused debugging.
Accordingly, \approach uses static analysis to decompose projects into \textit{field} and \textit{method} fragments (modifiers/type/name/value, or signature/body, respectively; application or test). For each, it extracts location, code, dependencies, types, file paths, inheritance, imports, and annotations using Tree-sitter~\cite{treesitter} and the Java Call Graph~\cite{java-callgraph}, becoming $10\times$ faster than \alphatrans's CodeQL-based approach~\cite{codeql}.

\subsection{Context-aware Type Resolution}
\label{subsec:type-resolution}

Automated type resolution is complex~\cite{guizzo2024mutation,terekhov2000realities}, addressed previously via symbolic rule-based approaches~\cite{broggi2023interoperability,george2010programming,lano2024using,roziere2020unsupervised,coco2018jpt,qiu1999programming} and, more recently, LLMs~\cite{jiang2025typybench,wang2024dainfer,peng2023generative}. In code translation specifically, prior work resolves types either purely via LLM (before or during translation) or via neuro-symbolic rules with LLM-guided inference; accuracy here is critical, as incorrect types cause runtime errors that block validation~\cite{ibrahimzada2025alphatrans,zhang2024scalable}.

\begin{figure}[t]
    \centering
    \includegraphics[width=0.9\columnwidth]{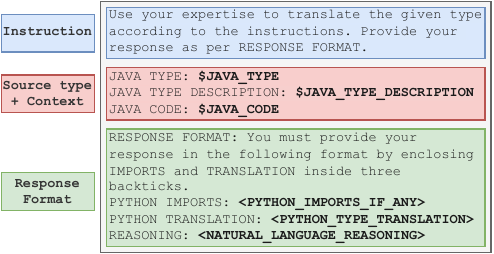}
    \caption{Context-aware type resolution prompt}
    \label{fig:ctr-prompt-template}
\end{figure}

Algorithm~\ref{alg:type-resolution} shows \approach's RAG-based, context-aware type resolution workflow. \approach extracts all source-language types (line~2); application-level types resolve trivially. For library types, it crawls source API documentation, extracts the type's usage context, and prompts the LLM (Figure~\ref{fig:ctr-prompt-template}) to produce target imports, type translation, and natural-language reasoning (lines 3--6).
To guard against hallucinated types, \approach validates the generated imports/translation by generating and executing a Python file that imports and references the type, checking only syntactic/runtime validity. Unresolved translations are inferred from other projects (lines 7--8); if none succeed, the type defaults to \texttt{\small{object}}. All translated types and their reasoning are stored in a \textit{Context-aware Type Map} for downstream use (line 9).

Proposed packages are installed at their latest available versions, since pinned mappings drift over time; any version or API mismatch surfaces through the executable import checks or as failing mock tests---at worst false negatives, never silent false positives. When no semantically equivalent Python library exists, the LLM is encouraged to implement the required logic itself (e.g., Python offers no peekable-iterator equivalent, so the model synthesizes one). Validation never depends on finding an equivalent library: third-party and user-defined objects are serialized by field-level recursive decomposition down to primitives and standard-library types (\S\ref{subsubsec:serialization}).

\sloppy Pre-translation resolution with shared reasoning follows from two findings: type-aware pre-conditioning reduces hallucinated/ill-typed API usage versus on-the-fly inference~\cite{jain2025mitigating,wang2025apirat}, and context-rationalized (not raw) type information improves LLM performance~\cite{jiang2025typybench,venkatesh2024emergence}---together improving translation faithfulness and consistency.
The same source type can map to different targets by context: Figure~\ref{listing:type} shows \emph{commons-cli}'s \texttt{\small{Options}} class, where both \texttt{\small{getMatchingOptions()}} and \texttt{\small{getRequiredOptions()}} return a Java \texttt{\small{List}}, but the former builds a modifiable list (warranting Python \texttt{\small{list}}), while the latter wraps its result in \texttt{\small{Collections.unmodifiableList}} (making \texttt{\small{list}} typing wrongly mutable). \approach's RAG-based resolution correctly infers a \texttt{\small{tuple}} for the latter.

\begin{figure}[t]
\lstset{
  escapeinside={(*@}{@*)},
  basicstyle=\scriptsize\ttfamily,
  columns=fullflexible,
  showlines=true
}
\begin{lstlisting}[language=Java]
(*@\texttt{------------- Java List mapped to Python list -------------}@*)
public List<String> getMatchingOptions(String opt) {
    final List<String> matchingOpts = new ArrayList<>();
    for (final String longOpt : longOpts.keySet()) {
        if (longOpt.startsWith(opt)) {
            matchingOpts.add(longOpt);
        }
    }
    return matchingOpts;
}
(*@\texttt{------------ Java List mapped to Python tuple ------------}@*)
public List getRequiredOptions() {
    return Collections.unmodifiableList(requiredOpts);
}
\end{lstlisting}
\vspace{-5pt}
\caption{\approach's different resolved Python types for \texttt{List} depending on the context}
\label{listing:type}
\end{figure}

Since fully automated resolution (unlike the manual intervention required in \alphatrans) cannot guarantee strict semantic equivalence, \approach's checks only ensure syntactic/runtime validity. It preserves inferred types in method signatures for structural consistency but does not strictly enforce them in the implementation; instead, predicted types and LLM reasoning serve as contextual guidance, letting the LLM refine type choices based on execution feedback.

\subsection{Translation with Integrated Isolated Validation}
\label{subsec:compositional-translation-validation}

\approach removes back edges from the project's call graph to make it acyclic, then translates method fragments in reverse topological order. Figure~\ref{fig:mocking} and Algorithm~\ref{alg:mocking-validation} outline the validation workflow: AspectJ-generated execution logs yield I/O behavior and side effects for each invocation in a Java unit test's call chain (lines~3--4), from which \approach builds a Python test calling the translated focal method directly (lines~17--18) while mocking all dependent application methods (lines~6--16). Each mock simulates the original method's return values and side effects (static field updates, exceptions). The test reconstructs initial state, invokes the focal method, and verifies return values, instance/parameter mutations, and global state changes via recursive equality checks (lines~19--26), relying on custom Java object serialization (supporting both standard-library and application types) for reproducible, isolated functional-equivalence validation.

\begin{algorithm}[t]
\scriptsize
\input{Resources/Algorithms/mocking-validation}
\end{algorithm}

\subsubsection{AspectJ-Based Runtime Interception} An AspectJ instrumentation layer intercepts public method calls, recording IO pairs, side effects, instance-state changes, and static-field modifications, using a broad pointcut over repository packages to exclude synthetic/library calls. On entry, it deep-copies the inputs (via \texttt{\small{CustomToStringConverter}}), snapshots \texttt{this} for instance methods, and reflects on static fields.

After execution, a second snapshot (parameters, return/exception, receiver state, global state) is compared against the first to detect mutations, serialized into a trace structure for behavior reconstruction and Python test generation. Executions proceed normally post-logging (exceptions re-thrown after recording), and the non-invasive advice layer integrates into Maven test runs without source changes, supporting whole-program or selective instrumentation (Figure~\ref{listing:runtime-interception}).

\begin{figure}[t]
\begin{lstlisting}[language = Python, columns=fullflexible, basicstyle=\scriptsize\ttfamily,showlines=true]
byte_array_json = json_obj.get("byte_array", None)
# similar routine to deserialize primitive arrays
byte_array = convert_to_python(byte_array_json)
byte_buffer = bytes([x & 0xFF for x in byte_array])
bytes_io_object = BytesIO(byte_buffer)
bytes_io_object.seek(int(json_obj.get("position", 0)))
\end{lstlisting}
\vspace{-5pt}
\caption{Deserialization of Java \texttt{\small{ByteArrayInputStream}} into Python \texttt{\small{io.BytesIO}}}
\label{listing:bytesio-deserialization}
\end{figure}

\subsubsection{Custom Serialization Workflow}\label{subsubsec:serialization} A \texttt{\small{CustomToString Converter}} class serializes Java objects to JSON needed to reconstruct equivalent Python objects, distinguishing application types (fine-grained introspection) from library classes (grouped handling, e.g., uniform treatment of \texttt{\small{Map}} implementations) by package prefix.
For application classes, the converter introspects fields across the class hierarchy, distinguishing static from instance fields, skipping synthetic/transient members, and recording the declaring class to resolve naming ambiguity. Edge cases get tailored handling: I/O subclasses (e.g., \texttt{\small{BufferedReader}}, \texttt{\small{FilterInputStream}}) account for both inherited and custom fields, while inner-class iterators serialize outer references (via synthetic \texttt{\small{this\$0}}) with cached identity tracking, ensuring cycle-free serialization even for deeply nested structures.

This custom serialization enables direct construction of target-language objects, eliminating the need for cross-language interoperability. Unlike \alphatrans, which relies on manually generated, class-specific glue code with fixed type mappings, our workflow is general-purpose and project-agnostic. \approach supports the Java SE 8 standard library for scalability, leveraging interface-level abstractions to generalize without handling every class individually, e.g., any \texttt{\small{java.util.List}} implementation (\texttt{\small{ArrayList}}, \texttt{\small{LinkedList}}, \texttt{\small{Vector}}) is handled uniformly via \texttt{\small{foreach}} iteration, with the Python deserializer reconstructing the appropriate collection type (per RAG-based type mapping). Third-party or user-defined classes undergo field-level recursive serialization down to primitive or standard-library types. For platform-dependent types (e.g., \texttt{\small{java.lang.Thread}}), \approach avoids overfitting to language-specific runtime states, instead only checking that the translated function returns a corresponding Python \texttt{\small{thread}}.

\begin{figure}[t]
\begin{lstlisting}[language = Java, columns=fullflexible, basicstyle=\scriptsize\ttfamily,showlines=true]
@Aspect
class LoggingAspect {
    @Pointcut("project-level main methods") void appPoints() {}
    @Around("appPoints()")
    Object log(ProceedingJoinPoint jp) {
        if (isTestOrSynthetic(jp)) return jp.proceed();
        updateJUnitContext();
        logStart(jp);
        recordInstanceAndArgs(jp);
        recordStaticFieldsBefore();
        Object result = null;
        try { result = jp.proceed(); }
        catch (Exception e) { logException(e); throw e; }
        recordStaticFieldsAfter();
        recordResults(jp, result);
        logEnd(jp);
        return result;
    }
    // helpers: detect JUnit methods, capture static
    // fields, find test method in stack, write logs, ...
}
\end{lstlisting}
\vspace{-8pt}
\caption{\approach interception/serialization process}
\vspace{-10pt}
\label{listing:runtime-interception}
\end{figure}

\begin{figure}[t]
\begin{lstlisting}[language = Java, columns=fullflexible, basicstyle=\scriptsize\ttfamily,showlines=true]
private static JSONObject handleByteArrayInputStream(
        ByteArrayInputStream bis, ...) {
    JSONObject bisDetails = new JSONObject();
    bisDetails.put("type", getTypeName(bis, ...));
    Class<?> bisClass = bis.getClass();
    /* Retrieve underlying byte array */
    Field bufField = bisClass.getDeclaredField("buf");
    bufField.setAccessible(true);
    byte[] array = (byte[]) bufField.get(bis);
    // similar method to handle primitive arrays
    bisDetails.put("byte_array",
        handlePrimitiveArray(array, ...);
    /* Retrieve current position */
    Field posField = bisClass.getDeclaredField("pos");
    posField.setAccessible(true);
    int position = posField.getInt(bis);
    bisDetails.put("position", position);
    return bisDetails;
}
\end{lstlisting}
\vspace{-5pt}
\caption{Type handling of \texttt{\small{ByteArrayInputStream}}}
\vspace{-5pt}
\label{listing:serailization-workflow}
\end{figure}

\subsubsection{Deserialization and Equality Verification Workflow} \approach reconstructs Python objects from \texttt{\small{Custom ToStringConverter}}-serialized JSON logs via a custom recursive pipeline, resolving each structure's Python class (application classes map to preloaded translated classes; Java types map to standard-library equivalents per context-aware type resolution). Primitives convert directly; compounds (tuples, object fields) use type-specific logic. A global dictionary tracks instantiated objects by serialization-time memory address, preserving reference identity and supporting cyclic references. Figure~\ref{listing:bytesio-deserialization} shows this converting a Java \texttt{\small{ByteArrayInputStream}} (serialized per Figure~\ref{listing:runtime-interception}) into a Python \texttt{\small{io.BytesIO}}.

\begin{figure}[t]
\begin{lstlisting}[language = Python, columns=fullflexible, basicstyle=\scriptsize\ttfamily, showlines=true]
class MockTest(unittest.TestCase):
  @patch('focal_method') @patch('helper_1') 
  @patch('helper_2')
  def test_mocking(self, helper_2, helper_1, focal_method):
    # Replace non-focal invocation by mocks
    helper_1.side_effect = SideEffect(method_1_invocation_1, method_1_invocation_2, ...)
    helper_2_.side_effect = SideEffect(method_2_invocation_1, method_2_invocation_2, ...)
    # Set up static fields before method calls
    update_static_fields(json.loads(orig_static_fields_json))
    # construct a Python equivalent of all arguments and the
    # original instance of the Java method invocation
    method_args = list(convert_to_python(json.loads(args_initial_json)))
    instance = convert_to_python(json.loads(instance_initial_json))
    # call the focal method (call chain replaced by mocks)
    returned_obj = instance.focal_method(*method_args)
    # construct a Python equivalent of final Java instance 
    # and check if instance is correct after invocation
    instance_reference = convert_to_python(json.loads(instance_final_json))
    self.assertTrue(recursive_equal(instance, instance_reference))
    # check if side effects are correct
    self.assertTrue(side_effect_is_correct(json.loads(focal_method_side_effect)))
    # construct a Python equivalent of the Java argument list 
    # after method invocation and check if it matches Python
    # method invocation
    method_args_reference = convert_to_python(json.loads(args_final_json))
    self.assertTrue(recursive_equal(method_args, method_args_reference))
    # construct a Python equivalent of the Java returned 
    # object and check if it matches the object returned by 
    # the Python method invocation
    returned_obj_reference = convert_to_python(json.loads(returned_obj_final_json))
    self.assertTrue(recursive_equal(returned_obj, returned_obj_reference))
\end{lstlisting}
\vspace{-8pt}
\caption{A mock test generated by \approach}
\label{listing:mock-test-example}
\end{figure}

For custom classes, deserialization introspects static/instance fields, simulating Java visibility via Python name mangling (e.g., \texttt{\small{private int x}} in \texttt{\small{Foo}} $\rightarrow$ \texttt{\small{\_Foo\_\_x}}); types inheriting from Java standard library classes ignore Java-specific fields. Enums reconstruct via \texttt{\small{\_name\_}}/\texttt{\small{\_value\_}}; exceptions use \texttt{\small{\_\_new\_\_}} to bypass argument enforcement while preserving messages.
A recursive equality checker then performs deep semantic comparison, with type-specific logic for primitives, collections, enums, and I/O streams (e.g., \texttt{\small{BytesIO}}/\texttt{\small{StringIO}} via buffer content) and recursive \texttt{\small{\_\_dict\_\_}} comparison for custom objects. It normalizes numerical tolerances (e.g., \texttt{\small{timedelta}} precision loss) and string-format differences (e.g., \texttt{\small{HashMap}} vs.\ \texttt{\small{dict}}), ensuring accurate behavioral equivalence while preserving identity, mutability, and visibility semantics.

\subsubsection{Automated Test Method Generation and Mocking} \approach generates mock tests validating each invoked method's translated Python counterpart against recorded Java behavior, in four steps. At \textbf{Reading the AspectJ Log} step, \approach retrieves the focal method's I/O/side-effect data (for comparison) and callees' data (for mock construction), isolating validation. At \textbf{Mock Decorator Insertion} step, it inserts decorators distinguishing constructors (\texttt{\small{<init>}}) from methods, patching each once; the focal method runs directly while others are mocked (simulating output, static-field updates, argument changes, exceptions). \textbf{Both public and private methods} are mockable: private methods (e.g., \texttt{\small{foo}} in \texttt{\small{Bar}}) become dunder-named (\texttt{\small{\_Bar\_\_foo}}), patchable via \texttt{\small{unittest.mock}}, avoiding the usual cross-language visibility limitation. At the \textbf{Method Body Construction} step, \approach initializes static fields/instances/parameters to mirror pre-invocation state, invokes the focal method (dependencies mocked), then verifies instance/parameter mutations, static-field changes, return values (recursive comparison), and exceptions against original behavior. Finally, at \textbf{File Writing} step, it assembles decorators, parameters, and bodies into a new test file per focal method, enabling isolated verification; runtime object setup/equivalence uses the deserialization workflow above.
For bodyless AMFs (interface, abstract, and empty-body methods), \approach additionally emits one \emph{static} mock test per method from the schema alone, checking that the translated module imports, the class defines the method with matching arity, and the stub stays trivial. Figure~\ref{listing:mock-test-example} shows an example mock test. All generated mock tests are publicly available~\cite{website}.

%% file: Resources/Algorithms/type-resolution.tex
\SetKwInOut{Input}{Input}
\SetKwInOut{Output}{Output}

\caption{Context-aware Type Resolution}
\label{alg:type-resolution}

\KwInputs{Subject Projects $Ps$}
\KwOutput{Context-aware Type Map $\mathit{CTM}$}

\ForEach{$\mathit{project} \in Ps$}{
    $\mathit{types} \gets \text{getProjectTypes}(\mathit{project})$\;
    \ForEach{$\mathit{type} \in \mathit{types}$}{
        $\mathit{doc} \gets \text{crawlTypeDocumentation}(\mathit{type})$\;
        $\mathit{code} \gets \text{getTypeCodeSnippet}(\mathit{type})$\;
        $\mathit{translation} \gets \text{translate\&Validate}(\mathit{type}, \mathit{doc}, \mathit{code})$\;
        \If{unsuccessful(translation)}{
            $\mathit{translation} \gets \text{resolveGlobally}(\mathit{type})$\;
        }
        $\mathit{CTM}[\mathit{project}][\mathit{type}] \gets \mathit{translation}$\;
    }
}

\Return{$\mathit{CTM}$}

%% file: Resources/Algorithms/mocking-validation.tex
\SetKwInOut{KwInput}{Input}
\SetKwInOut{KwOutput}{Output}

\caption{Mock-Based Isolated Validation}
\label{alg:mocking-validation}

\KwInput{Java Unit Tests $J$, Application Method Set $R$}
\KwOutput{Mock-Based Python Test Suite $T$}

$T \gets [\ ]$\;

\ForEach{$\mathit{test} \in J$}{
    $L \gets \text{runWithAspectJandCustomSerializer}(\mathit{test})$\;
    $\mathit{focal\_methods} \gets \text{extractAllInvocations}(L)$\;

    \ForEach{$\mathit{focal} \in \mathit{focal\_methods}$}{
        $C \gets \text{getDirectCallees}(\mathit{focal}, R)$\;

        $\mathit{test\_case} \gets \text{initTestFile}(\mathit{focal})$\;

        \tcp{\scriptsize Mock direct callees of the focal method}
        \ForEach{$c \in C$}{
            $M_c \gets \{$ \\
            \Indp
            $\text{output}_c,$ \\
            $\text{exception}_c,$ \\
            $\text{paramModifications}_c,$ \\
            $\text{instanceModifications}_c$ \\
            $\text{staticFieldsModifications}_c$ \\
            \Indm
            $\}$\;
            \text{injectMock}($\mathit{test\_case}, c, M_c$)\;
        }

        \tcp{\scriptsize Set up input of the focal method}
        $S_{init} \gets \{ \text{staticFields}, \text{focalInstance}, \text{focalParams} \}$\;
        \text{injectCall}($\mathit{test\_case}, \mathit{focal}, S_{init}$)\;

        \tcp{\scriptsize Verify focal method's output \& side effects}
        $\mathit{Expected} \gets \{$ \\
        \Indp
        $\text{output}_{\mathit{focal}},$ \\
        $\text{exception}_{\mathit{focal}},$ \\
        $\text{paramModifications}_{\mathit{focal}},$ \\
        $\text{instanceModifications}_{\mathit{focal}},$ \\
        $\text{staticFieldsModifications}_{\mathit{focal}}$ \\
        \Indm
        $\}$\;
        \text{injectVerifications}($\mathit{test\_case}, \mathit{Expected}$)\;

        \text{append}($T, \mathit{test\_case}$)\;
    }
}

\Return{$T$}

%% file: Resources/Sections/5-Evaluation.tex
\section{Evaluation}
\label{sec:evaluation}

To evaluate different aspects of \approach, we study the following research questions:

\vspace{3pt}
\begin{enumerate}[leftmargin=*,label=\bfseries RQ\arabic*:,nosep,wide=0pt]
  \item \textbf{Effectiveness.} To what extent can \approach effectively translate real-world projects? Does it improve functional equivalence compared to \alphatrans?

\vspace{3pt}
  \item \textbf{Reliability of Mock-based Isolated Validation.} How reliable is mock-based translation validation compared to GraalVM-based validation of \alphatrans and unit test generation and translation of MatchfixAgent?

\vspace{3pt}
  \item \textbf{Generalizability of \approach to Other Models.} How does \approach perform when integrated with different LLMs?

\vspace{3pt}
  \item \textbf{Contribution of Core Components.} Do \approach’s two key innovations, RAG-based type resolution and mock-based isolated validation, each play a significant role in improving its overall performance?

\end{enumerate}

\begin{table}
    \small
    \centering
    \caption{Effectiveness of \approach (\textbf{MB}) in repository-level code translation compared to \alphatrans (\textbf{AT}). \textbf{AMF}: \#Application Method Fragments, \textbf{Non-Validatble}: Fragments where isolated validation cannot validate (no test coverage for both approaches and GraalVM error for \alphatrans), \textbf{Success/Failure}: Fragments where isolated validation pass/fail (GraalVM for \alphatrans and mocking for \approach).}
    \footnotesize
    \setlength{\tabcolsep}{5pt}
    \renewcommand{\arraystretch}{0.95}
    \input{Resources/Tables/rq1-effectiveness}
    \label{table:rq1-effectiveness}
\end{table}

We use the \alphatrans benchmark of ten real-world Java projects, chosen for their diversity and relevance to repository-level translation challenges, particularly mock-based isolated validation.
\approach integrates Tree-sitter~\cite{treesitter} and Java Call Graph~\cite{java-callgraph} for static analysis, and Python Mock $5.2.0$~\cite{mockpy}, JUnit $4$/$5$~\cite{junit}, Pytest $8.2.1$~\cite{pytest}, JaCoCo~\cite{jacoco}, and Python's coverage tool~\cite{coverage-lib} for dynamic analysis. We use DeepSeekCoder-33B-Instruct~\cite{guo2024deepseek} as the primary LLM (matching AlphaTrans, for reproducibility and comparison), and Devstral-24B~\cite{devstral} for ablations showing generalizability.
All models use temperature 0 (deterministic output) with default generation parameters otherwise. Following \alphatrans's configuration, \approach also performs up to four re-prompting attempts, shared between syntax-based re-prompting (triggered by syntax errors) and feedback-based re-prompting (triggered by mock test failures).

\input{Resources/Sections/5-1-RQ1}

\input{Resources/Sections/5-2-RQ2}
\input{Resources/Sections/5-3-RQ3}
\input{Resources/Sections/5-4-RQ4}

%% file: Resources/Tables/rq1-effectiveness.tex
\begin{tabular}{l r cc cc cc}
\toprule
\multirow{2}{*}{\textbf{Subject}} &
\multirow{2}{*}{\textbf{AMF}} &
\multicolumn{2}{c}{\textbf{Non-Validatble (\%)}} &
\multicolumn{2}{c}{\textbf{Success (\%)}} &
\multicolumn{2}{c}{\textbf{Failure (\%)}} \\
\cmidrule(lr){3-4}\cmidrule(lr){5-6}\cmidrule(lr){7-8}
& & \textbf{AT} & \textbf{MB} & \textbf{AT} & \textbf{MB} & \textbf{AT} & \textbf{MB} \\
\midrule
cli         & 273 & 17.6 & 0.7 & 70.7 & 77.3 & 11.7 & 22.0 \\
codec       & 680 & 29.1 & 8.4 & 38.4 & 51.8 & 32.5 & 39.9 \\
csv         & 235 & 34.5 & 3.4 & 38.7 & 60.9 & 26.8 & 35.7 \\
exec        & 248 & 64.5 & 10.9 & 33.5 & 69.0 & 2.0 & 20.2 \\
fast-pfor   & 748 & 63.8 & 27.0 & 12.0 & 36.0 & 24.2 & 37.0 \\
fileupload  & 192 & 90.1 & 14.6 & 8.9 & 73.4 & 1.0 & 12.0 \\
graph       & 530 & 52.3 & 18.1 & 24.8 & 43.4 & 22.9 & 38.5 \\
jansi       & 409 & 80.4 & 26.2 & 8.1 & 48.4 & 11.5 & 25.4 \\
pool        & 682 & 92.7 & 34.6 & 6.0 & 53.4 & 1.3 & 12.0 \\
validator   & 646 & 58.4 & 2.6 & 30.5 & 57.6 & 11.2 & 39.8 \\
\midrule
\textbf{Total} & 4643 & 59.3 & 16.8 & \textbf{24.5} & \textbf{52.8} & 16.2 & 30.4 \\
\bottomrule
\end{tabular}

%% file: Resources/Sections/5-1-RQ1.tex
\subsection{RQ1: Effectiveness of \approach}
\label{subsec:rq-effectiveness}

\begin{table}[t]
    \centering
    \footnotesize
    \setlength{\tabcolsep}{5pt}
    \renewcommand{\arraystretch}{0.95}
    \caption{Line and branch coverage with existing developer-written tests vs.\ \approach's EvoSuite-augmented test suite. $\Delta$ is the absolute percentage-point gain.}
    \input{Resources/Tables/rq1-coverage}
    \label{table:rq1-coverage}
\end{table}

Table~\ref{table:rq1-effectiveness} compares \approach's effectiveness to \alphatrans. \approach evaluates functional equivalence for 83.2\% of AMFs across all 10 projects versus 40.73\% for \alphatrans, mainly due to \approach's EvoSuite generation and static mock tests for bodyless methods. We exclude LLM-generated tests to preserve determinism. EvoSuite lifts average line coverage from 64.47\% to 84.15\% and branch coverage from 63.57\% to 81.07\% (Table~\ref{table:rq1-coverage}), with \emph{fileupload}, \emph{pool}, \emph{validator}, and \emph{jansi} gaining over 25 points. Among validatable AMFs, \textbf{63.4\%} pass all mock tests versus \textbf{43.30\%} for \alphatrans/GraalVM.

Overall, \approach achieves \textbf{52.8\%} functional equivalence versus \alphatrans's \textbf{24.5\%}~\cite{guo2024deepseek}\footnote{\alphatrans reports 25.14\%, the union of GraalVM success (24.50\%) and test-translation success (2.88\%).}, driven by evaluating more AMFs, better type resolution, and more rigorous validation (ablation in \S\ref{subsec:ablation}).
Normalizing by validatable AMFs, \alphatrans has 1,891 validatable AMFs with a \textbf{39.87\%} failure rate, versus \approach's 3,863 validatable AMFs at \textbf{36.6\%}, lower despite validating $2.04\times$ more fragments. This gap widens once GraalVM's false positives (\S\ref{subsec:rq-mocking-validation}) and stronger-LLM effects (\S\ref{subsec:ablation}) are accounted for.

All translated AMFs scored 10/10 on Black~\cite{black}, indicating idiomatic output. Table~\ref{table:rq1-test_translation} shows \approach also produces more passing translated tests despite test translation being less isolated and prone to false positives. Translated tests appear only here and in RQ3's ATP comparison (Table~\ref{table:rq3-comparative}), to compare with \alphatrans on its own (weaker) metric; they are not part of \approach's validation pipeline.

\vspace{-8pt}
\mybox{\textnormal{\textbf{Summary.} \approach evaluates and validates functional equivalence of significantly higher AMFs than \alphatrans, 52.8\% compared to 24.5\%.}}

\subsubsection{More Validatble Fragments}

\begin{table}
    \centering
    \scriptsize
    \setlength{\tabcolsep}{3.4pt}
    \renewcommand{\arraystretch}{0.9}
    \caption{Effectiveness of \approach (\textbf{MB}) in validating AMFs via direct test translation, compared to \alphatrans (\textbf{AT}). \textbf{ATP}: Fragments All Test Pass, \textbf{OTF}: Fragments One Test Fail, \textbf{MTF}: Fragments Many Test Fail, \textbf{ATF}: Fragments All Test Fail, \textbf{TPR}: Test Pass Rate (all translated tests).}
    \input{Resources/Tables/rq1-test_translation}
    \label{table:rq1-test_translation}
\end{table}

\begin{figure}[t]
\vspace{-10pt}
\begin{lstlisting}[language = Java, columns=fullflexible, basicstyle=\scriptsize\ttfamily,showlines=true]
private final Map<String, Option> optionMap
        = new LinkedHashMap<>();
public Collection<Option> getOptions() {
    return optionMap.values();
}
\end{lstlisting}
\vspace{-8pt}
\caption{An example of GraalVM Error}
\label{listing:graal-error-1}
\end{figure}

GraalVM-based validation is limited in two ways: across the ten subject projects, 43.43\% of AMFs lack unit test coverage, and another 24.50\% cannot be validated by GraalVM despite substantial glue-code effort, due to complex types. Uncovered AMFs need automated test generation, but new tests often introduce unseen types. GraalVM's type-specific glue code requires regenerated, manually adapted bridge code for any type variation (e.g., generics, container element types). Even covered AMFs may trigger unresolvable ``GraalVM Errors.''
Figure~\ref{listing:graal-error-1} illustrates a host–guest mismatch: \texttt{\small{optionMap.values()}} returns a \texttt{\small{Collection<Option>}} view backed by a host-side \texttt{\small{LinkedHashMap}}, but GraalPython lacks interop support for \texttt{\small{java.util.Collection}} containing unregistered objects like \texttt{\small{Option}}. \approach's mocking workflow instead intercepts the call at runtime, identifies the \texttt{\small{LinkedHashMap}} type, and constructs a faithful Python equivalent.

\mybox{\textnormal{\textbf{Summary.} Mock-based isolated validation scales the number of validatable AMFs by overcoming limitations of GraalVM.}}

\subsubsection{Automated and More Reliable, Context-Aware Type Mapping}

\alphatrans reports 91.99\% type translation accuracy versus \approach's 89.00\%---a slight, expected drop from \approach's richer context prompting occasionally increasing hallucinations (especially imports). The key difference is scalability: \alphatrans needs manual intervention for 8.01\% of failures plus correction of 12.24\% of LLM-based translations (55 developer-hours), while \approach assigns \texttt{\small{object}} to unresolved types with no manual effort, matching accuracy.
\alphatrans's fixed mapping can also mislead despite passing runtime checks: in \texttt{\small{commons-csv}}, \texttt{\small{java.nio.CharBuffer}} maps to \texttt{\small{list[str]}}, but Python lists lack buffer semantics and can't be passed to \texttt{\small{TextIOWrapper.readinto()}}; \approach instead infers \texttt{\small{bytearray}}. Similarly (Figure~\ref{listing:context-aware-benefit}), Java's \texttt{\small{Appendable}}--mapped by \alphatrans to a context-free \texttt{\small{List}}---is more plausibly a \texttt{\small{StringBuilder}} or \texttt{\small{PrintWriter}} given surrounding delimiter-handling code. \approach correctly suggests \texttt{\small{io.StringIO}} in both cases, while \alphatrans's fixed mapping yields erroneous calls like \texttt{\small{StringIO("".join(appendable))}}.

\mybox{\textnormal{\textbf{Summary.} Context-aware type resolution eliminates the need for manual effort by \alphatrans, enabling more faithful handling of complex types and preserving semantics during code translation.}}

\begin{figure}[t]
\begin{lstlisting}[language = Java, columns=fullflexible, basicstyle=\scriptsize\ttfamily,showlines=true]
public synchronized void println(final Appendable appendable)
        throws IOException {
    if (getTrailingDelimiter()) {
        append1(getDelimiterString(), appendable);
    }
    if (recordSeparator != null) {
        append1(recordSeparator, appendable);
    }
}
\end{lstlisting}
\vspace{-8pt}
\caption{An example showcasing effectiveness of context-aware type resolution}
\label{listing:context-aware-benefit}
\end{figure}

\begin{figure}[t]
\begin{lstlisting}[language=Java, aboveskip=10pt,belowskip=0pt]
(*@\texttt{------------------- Java Code -------------------}@*)
if (it.hasNext()) {
    buf.append(", ");
}
\end{lstlisting}
\begin{lstlisting}[language=Python, aboveskip=0pt,belowskip=10pt]
(*@\texttt{---------------- LLM Translation ----------------}@*)
# superfluously advances the iterator
if next(it, None) is not None:
    buf.write(", ")
\end{lstlisting}
\vspace{-8pt}
\caption{GraalVM false positive due to API mismatch}
\label{listing:graal-fp-1}
\end{figure}

%% file: Resources/Tables/rq1-coverage.tex
\begin{tabular}{l cc cc cc}
\toprule
\multirow{2}{*}{\textbf{Subject}} &
\multicolumn{2}{c}{\textbf{Existing Tests (\%)}} &
\multicolumn{2}{c}{\textbf{w/ EvoSuite (\%)}} &
\multicolumn{2}{c}{\textbf{$\Delta$ (pp)}} \\
\cmidrule(lr){2-3}\cmidrule(lr){4-5}\cmidrule(lr){6-7}
& \textbf{Line} & \textbf{Branch} & \textbf{Line} & \textbf{Branch} & \textbf{Line} & \textbf{Branch} \\
\midrule
cli         & 96.25 & 93.99 & 99.04 & 97.13 & +2.79  & +3.14  \\
codec       & 91.09 & 92.19 & 98.43 & 96.80 & +7.34  & +4.61  \\
csv         & 88.26 & 81.47 & 93.94 & 89.37 & +5.68  & +7.90  \\
exec        & 61.70 & 50.67 & 82.05 & 75.56 & +20.35 & +24.89 \\
fast-pfor   & 66.09 & 65.49 & 68.90 & 71.77 & +2.81  & +6.28  \\
fileupload  & 36.26 & 44.85 & 85.55 & 83.33 & +49.29 & +38.48 \\
jansi       & 23.68 & 18.38 & 54.83 & 43.63 & +31.15 & +25.25 \\
graph       & 84.98 & 78.54 & 98.23 & 95.38 & +13.25 & +16.84 \\
pool        & 33.94 & 44.44 & 66.28 & 65.08 & +32.34 & +20.64 \\
validator   & 62.45 & 65.67 & 94.29 & 92.64 & +31.84 & +26.97 \\
\midrule
\textbf{Total} & \textbf{64.47} & \textbf{63.57} & \textbf{84.15} & \textbf{81.07} & \textbf{+19.68} & \textbf{+17.50} \\
\bottomrule
\end{tabular}

%% file: Resources/Tables/rq1-test_translation.tex
\begin{tabular}{l r cc cc cc cc cc}
\toprule
\multirow{2}{*}{\textbf{Subject}} &
\multirow{2}{*}{\textbf{AMF}} &
\multicolumn{2}{c}{\textbf{ATP}} &
\multicolumn{2}{c}{\textbf{OTF}} &
\multicolumn{2}{c}{\textbf{MTF}} &
\multicolumn{2}{c}{\textbf{ATF}} &
\multicolumn{2}{c}{\textbf{TPR}} \\
\cmidrule(lr){3-4}\cmidrule(lr){5-6}\cmidrule(lr){7-8}\cmidrule(lr){9-10}\cmidrule(lr){11-12}
& & \textbf{AT} & \textbf{MB} & \textbf{AT} & \textbf{MB} & \textbf{AT} & \textbf{MB} & \textbf{AT} & \textbf{MB} & \textbf{AT} & \textbf{MB} \\
\midrule
cli        & 273 & 8.4 & 6.2 & 10.6 & 11.4 & 32.2 & 12.5 & 7.0 & 18.3 & 10.1 & 20.2 \\
codec      & 680 & 4.1 & 4.1 & 4.4 & 11.3 & 12.8 & 28.7 & 4.1 & 22.2 & 9.4 & 20.4 \\
csv        & 235 & 0.0 & 1.3 & 13.6 & 3.0 & 0.0 & 0.9 & 2.6 & 4.3 & 0.0 & 1.2 \\
exec       & 248 & 4.4 & 10.9 & 2.0 & 8.9 & 7.3 & 6.9 & 6.9 & 13.3 & 19.3 & 48.1 \\
fast-pfor  & 748 & 4.3 & 21.7 & 1.7 & 10.2 & 3.7 & 11.2 & 3.1 & 8.8 & 20.1 & 44.7 \\
fileupload & 192 & 3.7 & 13.0 & 6.8 & 6.3 & 1.0 & 12.5 & 0.0 & 2.6 & 63.4 & 48.0 \\
graph      & 530 & 0.0 & 11.3 & 0.9 & 15.7 & 0.2 & 9.4 & 0.6 & 10.0 & 11.0 & 34.7 \\
jansi      & 409 & 0.2 & 9.0 & 1.0 & 8.6 & 0.0 & 4.9 & 0.0 & 4.2 & 1.1 & 28.2 \\
pool       & 682 & 1.6 & 1.5 & 1.0 & 1.5 & 0.3 & 0.4 & 0.0 & 0.7 & 6.6 & 53.3 \\
validator  & 646 & 3.3 & 4.6 & 5.4 & 2.5 & 4.2 & 0.9 & 4.0 & 2.9 & 11.7 & 60.3 \\
\midrule
\textbf{Total} & 4643 & \textbf{2.9} & \textbf{8.6} & 3.7 & 7.9 & 5.4 & 9.4 & 2.6 & 8.8 & \textbf{9.8} & \textbf{31.4} \\
\bottomrule
\end{tabular}

%% file: Resources/Sections/5-2-RQ2.tex
\subsection{RQ2: Reliability of Mock-based Isolated Validation}
\label{subsec:rq-mocking-validation}

RQ1 shows mock-based isolation validates more AMFs, raising the question of reliability: a false negative reports failure for an equivalent translation; a false positive (more problematic) reports success for a non-equivalent one. We manually investigated mock tests and \alphatrans's GraalVM results (plus its human-repair study) across three projects, analyzing translations and validation artifacts for 700 AMFs.
These 700 AMFs are \textbf{all} AMFs from \emph{cli}, \emph{csv}, and \emph{fileupload}, one project per API-diversity group among the ten benchmark projects: (1) parsing/structured logic (\emph{cli}, \emph{codec}, \emph{validator}, \emph{jansi}); (2) data structures/stream processing (\emph{csv}, \emph{graph}, \emph{pool}); (3) I/O- or OS-intensive workflows (\emph{fileupload}, \emph{fast-pfor}, \emph{exec}). Rather than randomly sampling 700 from 4,643 AMFs, risking project imbalance and reduced reliability, we analyze \textbf{all} fragments in call-graph order from the three chosen projects. Detailed logs are in the \approach artifact~\cite{website}.

Each AMF was independently labeled by \emph{three inspectors} (two authors, one external non-author), blinded to validation outcomes, with success requiring majority agreement (2/3). Inter-rater agreement was $\kappa = 0.79$ (Fleiss' kappa), substantial, with an ``almost perfect'' threshold of $0.81$.
An AMF is labeled \emph{success} iff it preserves the source fragment's observable behavior---return values, side effects (parameter/instance/static-state mutations), and exception behavior---modulo cross-language representation differences (e.g., \texttt{\small{HashMap}} vs.\ \texttt{\small{dict}} ordering, numeric precision). Inspectors had full access to the Java source and the Python translation, compared each fragment against the manually translated reference projects released in the \alphatrans artifact, and executed the translation end-to-end whenever possible. Differences stemming only from nondeterministic or environment-dependent values (system time, randomness, hash-based identity) were judged correct and attributed to the validation oracle.

\begin{table}
    \small
    \centering
    \caption{Manual analysis of translations and mock-based/GraalVM-based/human validation.}
    \input{Resources/Tables/rq2-manual}
    \label{table:rq2-manual}
\end{table}

\begin{figure}[t]
    \centering
    \includegraphics[width=0.6\columnwidth]{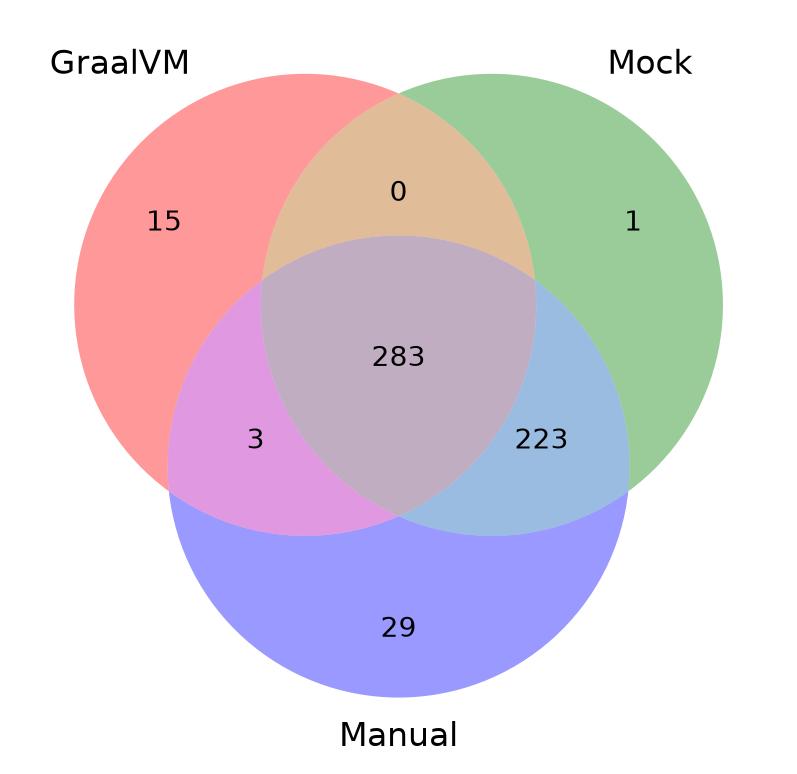}
    \caption{Success overlap across GraalVM-based, mock-based, and human validation}
    \label{fig:venn}
\end{figure}

Table~\ref{table:rq2-manual} shows mock-based validation achieving a substantially higher success rate than GraalVM across all three projects, mainly due to producing no errors on complex generic types and validating more AMFs via EvoSuite augmentation. Mocking shows a \textbf{single false positive (0.1\%)} and only \textbf{0.3\% false negatives}, versus GraalVM's \textbf{2.1\% false positives} and \textbf{14.0\% false negatives}, despite GraalVM validating fewer AMFs overall.
Figure~\ref{fig:venn} shows substantial agreement among all three methods (283 AMFs labeled "translation success" by GraalVM, mocking, and human validation alike). Against human ground truth, mocking validated 223 AMFs GraalVM missed, while GraalVM validated only three cases mocking didn't cover (15 GraalVM-only successes are later shown to be false positives). Both have unvalidatable cases, but GraalVM has notably more, highlighting its key limitation: dependence on original test-suite coverage.

\vspace{-5pt}
\mybox{\textnormal{\textbf{Summary.} Manual analysis of 700 AMFs from three representative projects shows mocking achieves a single false positive and only 0.3\% false negatives, substantially outperforming GraalVM (2.1\% false positives, 14.0\% false negatives). On the same projects, MatchFixAgent~\cite{ibrahimzada2026matchfixagent} reports an 11.0\% combined mislabel rate despite using a stronger LLM (Claude~Sonnet~3.7) than \approach's.} Mock-based validation is thus both more accurate and able to validate more fragments via automated test augmentation.}

\subsubsection{Comparison with MatchFixAgent}

MatchFixAgent~\cite{ibrahimzada2026matchfixagent}, the SOTA LLM-based in-isolation validator, drops language interoperability entirely, using a frontier LLM (Claude~Sonnet~3.7~\cite{anthropic2025sonnet37}) to judge equivalence directly from the source, candidate translation, and project context. This avoids GraalVM pitfalls but is more expensive and less reliable than mocking. MatchFixAgent reports a cost of \$1.22, on average, to validate an AMF (it assumes the translation exists, and only repairs if it deems it needed). MatchFixAgent translates only four out of ten \alphatrans subjects, i.e., 1{,}346 AMFs as reported. With the mentioned average cost, this estimates their cost to be around \$1{,}642, making it costly to scale compared to \approach's open-source LLMs (e.g., DeepSeekCoder-33B-Instruct~\cite{guo2024deepseek}, \S\ref{subsec:ablation}), which keeps costs compute-bound rather than per-token.

Reliability-wise, MatchFixAgent suffers from compounding errors due to LLM reasoning uncertainty and shallow test coverage. Of 758 jointly-validatable fragments across those four projects, 233 disagreed, with 62 confirmed in \alphatrans's favor, lower-bounding error at $62/758 = 8.18\%$. \emph{commons-validator} is structurally atypical, ${\sim}13$ near-identical, one-line validator classes wrapping logic in shared \texttt{Abstract*Validator} classes, contributing 458/758 ($60\%$) trivially-equivalent fragments. Restricting to \emph{cli}, \emph{csv}, and \emph{fileupload} (our manual study's projects, with substantive logic) reveals MatchFixAgent's $11.0\%$ combined mislabel rate (300 jointly validatable, 108 disagreements, 33 confirmed in \alphatrans's favor). \approach yields a single false positive and 0.3\% false negatives on the same projects (Table~\ref{table:rq2-manual}), substantially more reliable and cheaper despite MatchFixAgent's stronger model.

\begin{table}[t]
    \centering
    \footnotesize
    \setlength{\tabcolsep}{5pt}
    \renewcommand{\arraystretch}{0.95}
    \caption{Reported equivalence (\%) on MatchFixAgent's (Claude~3.7~Sonnet + Claude~Code) four projects vs.\ \approach's executed mock validation.}
    \input{Resources/Tables/rq2-matchfix}
    \label{table:rq2-matchfix}
\end{table}

Table~\ref{table:rq2-matchfix} contrasts per-project equivalence rates on MatchFixAgent's four subjects. \approach's rates are comparable while relying on smaller, older, open-source models, and its verdicts come from executed mock tests, whereas MatchFixAgent's are LLM judgments: its own manual investigation of disagreements with existing tools (its Table~1) rules 30\%, 30\%, 25\%, and 20\% of disagreements on \emph{cli}, \emph{csv}, \emph{fileupload}, and \emph{validator} in favor of the existing tool, and our larger 700-AMF manual study shows mocking has far fewer false positives and negatives. MatchFixAgent also relies on Claude~3.7~Sonnet with Claude~Code---a newer, stronger, and costlier frontier model than DeepSeekCoder-33B---at \$1.22 per verdict (${\sim}\$1{,}642$ for its 1,346 fragments).

\subsubsection{GraalVM False Positives}
Figure~\ref{listing:graal-fp-1} shows Java's \texttt{\small{hasNext()}} mistranslated as Python's \texttt{\small{next()}}. The latter erroneously advances the iterator. GraalVM passes since no test checks iterator state; mocking catches the unnecessary advancement.
GraalVM also cannot use EvoSuite tests, relying on developer tests that often miss edge cases. In Figure~\ref{listing:graal-fp-2}, an empty-string \texttt{\small{option}} should return \texttt{\small{longOption}} (non-\texttt{\small{null}}), but the Python translation wrongly takes the \texttt{\small{else}} branch, missed by developer tests, caught by EvoSuite mocks.

\begin{figure}[t]
\vspace{-10pt}
\begin{lstlisting}[language=Java, aboveskip=10pt,belowskip=0pt]
(*@\texttt{------------------- Java Code -------------------}@*)
return option == null ? longOption : option;
\end{lstlisting}
\begin{lstlisting}[language=Python, aboveskip=0pt,belowskip=10pt]
(*@\texttt{---------------- LLM Translation ----------------}@*)
# goes to "else" if option is empty string
return self.__option if self.__option else self.__longOption
\end{lstlisting}
\vspace{-8pt}
\caption{GraalVM false positive due to test incompleteness}
\label{listing:graal-fp-2}
\end{figure}

\subsubsection{GraalVM False Negatives}

GraalVM shows substantial false negatives: glue-code bugs often surface as runtime errors rather than compilation failures, misreporting correct translations as test failures. Figure~\ref{listing:polyglot-exception} shows a resulting \texttt{\small{PolyglotException}}: \texttt{\small{OptionBuilder.withArgName("ICSE")}} triggers a Python-to-Java field assignment (\texttt{\small{OptionBuilder.\_\_argName = name}}) on a private static \texttt{\small{String}}, but a latent glue bug fails to expose that field, so GraalVM can't assign the Python string, raising the exception. Exporting the field (\texttt{\small{@Export}} or \texttt{\small{allowHostAccess=ALL}}) would fix this. The error stems from glue-code misconfiguration, not a translation defect.

\begin{figure}[t]
\begin{lstlisting}[language=Java, aboveskip=10pt, belowskip=0pt]
(*@\texttt{------------------- Java Code -------------------}@*)
public static OptionBuilder withArgName(final String name) {
    OptionBuilder.argName = name;
    return INSTANCE;
}
\end{lstlisting}
\begin{lstlisting}[language=Python, aboveskip=0pt, belowskip=10pt]
(*@\texttt{---------------- LLM Translation ----------------}@*)
@staticmethod
def withArgName(name: str) -> "OptionBuilder":
    # Throws PolyglotException (bug in glue code)
    OptionBuilder.__argName = name
    return OptionBuilder.__INSTANCE
\end{lstlisting}
\vspace{-8pt}
\caption{A translation that results in GraalVM \texttt{\small{PolyglotException}}}
\vspace{-10pt}
\label{listing:polyglot-exception}
\end{figure}

\begin{figure}[t]
\begin{lstlisting}[language=Java, aboveskip=10pt, belowskip=0pt]
(*@\texttt{------------------- Java Code -------------------}@*)
private static void reset() {
    argCount = Option.UNINITIALIZED;
}
\end{lstlisting}
\begin{lstlisting}[language=Python, aboveskip=0pt, belowskip=10pt]
(*@\texttt{---------------- LLM Translation ----------------}@*)
@staticmethod
def __reset() -> None:
    import Option
    # throws NoClassDefFoundError (bug in glue code)
    OptionBuilder.__argCount = Option.UNINITIALIZED
\end{lstlisting}
\vspace{-8pt}
\caption{A translation that results in GraalVM Error (\texttt{\small{NoClassDefFoundError}})}
\label{listing:graal-error}
\end{figure}

Filtering GraalVM errors like \texttt{\small{PolyglotException}} as false alarms is tempting, but distinguishing translation bugs from glue-code bugs is often hard, especially for exceptions like \texttt{\small{NullPointerException}} or \texttt{\small{NoClassDefFoundError}}. Figure~\ref{listing:graal-error} shows glue code failing to expose class \texttt{\small{Option}} (including \texttt{\small{UNINITIALIZED}}) to the polyglot engine, causing \texttt{\small{NoClassDefFoundError}} despite the Python file correctly importing \texttt{\small{Option}}, since GraalVM's host--guest boundary means Python's \texttt{\small{import}} only affects the guest runtime, not the host Java classloader.
Even with all glue-code bugs fixed, GraalVM can still fail from inherent interoperability limits: Figure~\ref{listing:correct-translation-graal-error} (\emph{commons-cli}) shows correct code and glue logic still raising a runtime exception, since Java's \texttt{\small{contains()}} and \texttt{\small{remove(index)}} operate on indices rather than iterators.

\begin{figure}[t]
\begin{lstlisting}[language=Java, aboveskip=10pt, belowskip=0pt]
(*@\texttt{------------------- Java Code -------------------}@*)
if (requiredOpts.contains(key)) {
    requiredOpts.remove(requiredOpts.indexOf(key));
}
\end{lstlisting}
\begin{lstlisting}[language=Python, aboveskip=0pt, belowskip=10pt]
(*@\texttt{---------------- LLM Translation ----------------}@*)
if key in self.__requiredOpts:
    self.__requiredOpts.remove(key)
\end{lstlisting}
\vspace{-8pt}
\caption{A correct translation with Graal Error}
\vspace{-15pt}
\label{listing:correct-translation-graal-error}
\end{figure}

Under GraalVM, Python's \texttt{\small{in}} operator uses a \texttt{\small{java.util.Iterator}}, but \texttt{\small{remove(value)}} calls the collection's \texttt{\small{remove()}} rather than the iterator's, triggering \texttt{\small{java.util.ConcurrentModificationException}} due to Java's fail-fast iterator semantics. The translation is correct, as Python iterators lack modification counts, reflecting a semantic mismatch in GraalVM's interop layer, not a translation or glue-code flaw. Other false-negative sources include flaky tests from \alphatrans's assertion-level test decomposition, which introduces order dependencies~\cite{palomba2014testsmell,zhang2014independence} and spurious failures.

\subsubsection{Mocking False Positives}
We found a single false positive (Figure~\ref{listing:mbird-fp}): the translation buffers the usage text in a \texttt{\small{StringIO}} and prints it only after the delegate returns, while Java streams it to \texttt{\small{System.out}}. The divergence is confined to \emph{stdout}, which the mock oracle (checking return values, arguments, state mutations, and exceptions) does not observe. Such cases remain possible wherever side effects escape the recorded state, e.g., console or remote-server output.

\begin{figure}[t]
\begin{lstlisting}[language=Java, aboveskip=10pt, belowskip=0pt]
(*@\texttt{------------------- Java Code -------------------}@*)
final PrintWriter pw = new PrintWriter(System.out);
printHelp3(pw, width, cmdLineSyntax, /*...*/ autoUsage);
pw.flush();
\end{lstlisting}
\begin{lstlisting}[language=Python, aboveskip=0pt, belowskip=10pt]
(*@\texttt{---------------- LLM Translation ----------------}@*)
pw = io.StringIO()
self.printHelp3(pw, width, cmdLineSyntax, ..., autoUsage)
# stdout-only divergence: invisible to the mock oracle
print(pw.getvalue(), end='')
\end{lstlisting}
\vspace{-8pt}
\caption{The single mocking false positive (stdout-only)}
\label{listing:mbird-fp}
\end{figure}

\subsubsection{Mocking False Negatives}

\begin{figure}[t]
\begin{lstlisting}[language=Java, aboveskip=10pt, belowskip=0pt]
(*@\texttt{------------------- Java Code -------------------}@*)
@Override
public int hashCode() {
    return Objects.hash(longOption, option);
}
\end{lstlisting}
\begin{lstlisting}[language=Python, aboveskip=0pt, belowskip=10pt]
(*@\texttt{---------------- LLM Translation ----------------}@*)
def __hash__(self) -> int:
    return hash((self.longOption, self.option))
\end{lstlisting}
\vspace{-8pt}
\caption{\approach false negative example}
\label{listing:mbird-fn}
\end{figure}

Mock-based validation can yield rare false negatives when a function lacks deterministic I/O, yet mocking assumes identical outputs across invocations, common with system time, randomness, or hash-based identity. Figure~\ref{listing:mbird-fn} shows the mocking framework wrongly expecting Python's \texttt{hashCode()} equivalent to match a Java-observed integer, despite Python's differing, run-variable hash algorithm. The translation is correct; only validation fails to account for nondeterminism.
EvoSuite tests show similar flakiness when hard-coding environmental assumptions (locale, OS, system time), a broader issue with random test generation, not unique to mocking. Still, mock-based false negatives remain rarer than GraalVM's.

\vspace{-3pt}
\mybox{\textnormal{\textbf{Summary.} GraalVM-based validation suffers from false positives (mostly incomplete tests or API mismatches) and false negatives (mostly glue-code bugs, runtime interop errors, flaky tests). Mock-based validation exhibits few false negatives, primarily due to nondeterministic or environment-dependent behavior.}}
\vspace{-3pt}

%% file: Resources/Tables/rq2-manual.tex
\centering
\begin{tabular}{l@{\hskip 15pt}c@{\hskip 15pt}c@{\hskip 15pt}c}
\toprule
\textbf{Category} & \textbf{GraalVM} & \textbf{Mocking} & \textbf{Human} \\
\midrule
Success       & 301 (43.0\%) & 507 (72.4\%) & 538 (76.9\%) \\
Failure       & 97 (13.9\%)  & 153 (21.9\%) & 162 (23.1\%) \\
Error         & 102 (14.6\%) & 0 (0.0\%)    & - \\
Not Covered   & 200 (28.6\%) & 40 (5.7\%)   & - \\
\midrule
\textbf{Total} & 700 & 700 & 700  \\
\midrule
\multicolumn{1}{l}{\textbf{Mislabels}} & \textbf{GraalVM} & \textbf{Mocking} & - \\
\midrule
False Positive & \textbf{15} (2.1\%) & \textbf{1} (0.1\%) & - \\
False Negative & \textbf{98} (14.0\%) & \textbf{2} (0.3\%) & - \\
\bottomrule
\end{tabular}

%% file: Resources/Tables/rq2-matchfix.tex
\begin{tabular}{l cccc}
\toprule
\textbf{Validator} & \textbf{cli} & \textbf{csv} & \textbf{fileupload} & \textbf{validator} \\
\midrule
MatchFixAgent            & 76.9 & 78.7 & 75.0 & 74.8 \\
\approach{} (DS-33B)     & 77.3 & 60.9 & 73.4 & 57.6 \\
\approach{} (Devstral-24B) & 85.0 & 71.5 & 74.0 & 69.7 \\
\bottomrule
\end{tabular}

%% file: Resources/Sections/5-3-RQ3.tex
\subsection{RQ3: Generalizability of \approach to Other Models}
\label{subsec:rq-debugger}

\begin{table}[t]
    \centering
    \footnotesize
    \setlength{\tabcolsep}{5pt}
    \renewcommand{\arraystretch}{0.95}
    \caption{Comparative Analysis Between DeepSeekCoder-33B-Instruct and Devstral-24B concerning All Test Pass (\textbf{ATP}) and \textbf{Mock Success}.}
    \input{Resources/Tables/rq3-comparative}
    \label{table:rq3-comparative}
\end{table}

We investigate whether \approach's effectiveness is tied to a specific LLM (DeepSeekCoder-33B-Instruct~\cite{guo2024deepseek}) by replicating RQ1 with a smaller, more recent model, Devstral-24B~\cite{devstral}.
Table~\ref{table:rq3-comparative} shows Devstral-24B consistently scoring higher on most metrics, with comparable overall fidelity. \approach with Devstral-24B achieves \textbf{57.6\%} functional equivalence versus DeepSeekCoder-33B-Instruct's 52.8\%, outperforming it on mock success rate in nine of ten projects. More notably, Devstral-24B achieves a 14.34\% all-test-pass (ATP) rate for translated tests versus 8.59\% for DeepSeekCoder-33B-Instruct (nine of ten projects)---significant, since test translation, dependent on library semantics and implicit I/O/environment assumptions, is harder than source translation.

This performance on a medium-sized open-source model suggests \approach's effectiveness does not critically depend on the underlying LLM, consistent with its design: RAG-based in-context type resolution and mock-based isolated validation aren't optimized for any specific model. Type resolution is relatively lightweight (recognizing type semantics from context rather than deep reasoning), suiting smaller LLMs, while mock-based validation is LLM-independent.

\vspace{-3pt}
\mybox{\textnormal{\textbf{Summary.} \approach's context-aware type resolution and heuristic-based validation generalize well across different models.}}
\vspace{-8pt}

%% file: Resources/Tables/rq3-comparative.tex
\begin{tabular}{l cc cc}
\toprule
\multirow{2}{*}{\textbf{Subject}} &
\multicolumn{2}{c}{\textbf{DeepSeekCoder-33B}} &
\multicolumn{2}{c}{\textbf{Devstral-24B}} \\
\cmidrule(lr){2-3}\cmidrule(lr){4-5}
& \textbf{Mock Suc. (\%)} & \textbf{ATP (\%)} & \textbf{Mock Suc.(\%)} & \textbf{ATP (\%)} \\
\midrule
cli        & 77.29 & 6.23 & 84.98 & 8.79 \\
codec      & 51.76 & 4.12 & 61.03 & 10.44 \\
csv        & 60.85 & 1.28 & 71.49 & 5.96 \\
exec       & 68.95 & 10.89 & 72.58 & 10.08 \\
fast-pfor  & 35.96 & 21.66 & 37.57 & 25.94 \\
fileupload & 73.44 & 13.02 & 73.96 & 16.15 \\
graph      & 43.40 & 11.32 & 46.79 & 12.08 \\
jansi      & 48.41 & 9.05 & 51.34 & 13.69 \\
pool       & 53.37 & 1.47 & 50.88 & 2.35 \\
validator  & 57.59 & 4.64 & 69.66 & 26.47 \\
\midrule
\textbf{Total} & \textbf{52.79} & \textbf{8.59} & \textbf{57.57} & \textbf{14.34} \\
\bottomrule
\end{tabular}

%% file: Resources/Sections/5-4-RQ4.tex
\subsection{RQ4: Contribution of Core Components (Ablations)}
\label{subsec:ablation}

\subsubsection{Mocking}
We run two ablations. First, we remove all EvoSuite-generated tests to isolate test-augmentation's contribution to validating otherwise-uncovered AMFs. Second, we disable on-the-fly mock-based validation during translation, relying only on translated developer-written tests and syntactic/runtime checks, running mock-based validation just once after translation completes, testing whether coverage alone suffices. We exclude GraalVM from this ablation since it requires labor-intensive manual glue code for generated tests~\cite{ibrahimzada2025alphatrans}; RQ1 already shows mocking's superiority.

Columns 3--4 of Table~\ref{table:rq4-ablation} show both ablations substantially degrade performance. Without EvoSuite, equivalence drops from \textbf{52.8\%} to \textbf{35.0\%} (still above \alphatrans's \textbf{25.14\%}), mainly hurting projects whose developer tests leave most methods uncovered (\emph{fileupload}, \emph{fast-pfor}) while well-covered ones (\emph{cli}, \emph{csv}) stay stable: the coverage EvoSuite adds is an important driver of \approach's gains, though not the only one. Without on-the-fly mocking, equivalence drops to \textbf{44.8\%}, mock failures produce informative traces that localize the mismatch to specific runtime objects (return values, fields, arguments, exceptions), giving the refinement loop concrete debugging signal; without them, feedback degenerates into the coarse, often binary pass/fail verdicts that GraalVM-style validation provides. This confirms \approach's gains come from both coverage and precise per-invocation feedback, enabled by direct object construction and isolated mocking, unlike GraalVM's coarser validation.

\mybox{\textnormal{\textbf{Summary.} Mocking contributes to \approach{}'s performance both by supporting EvoSuite-based test augmentation and on-the-fly validation with meaningful feedback, demonstrating that increased coverage combined with precise, per-invocation feedback drives reliable and informative translation validation beyond what existing approaches provide.}}

\subsubsection{Context-aware Type Resolution}
Reverting to \alphatrans's fixed, manually curated type mappings drops equivalence from \textbf{52.8\%} to \textbf{48.0\%} (Column 5), a modest gap given \approach's resolution is fully automated versus \alphatrans's 55-human-hour, 360-type manual curation. \approach needs no manual intervention, defaulting ambiguous types to \texttt{object} while remaining scalable.


\begin{table}[t]
    \centering
    \footnotesize
    \setlength{\tabcolsep}{5pt}
    \renewcommand{\arraystretch}{0.95}
    \caption{Effectiveness of \approach (\textbf{MB}) components on validation results. \textbf{ES}: EvoSuite, \textbf{OTF}: Mocking On-the-fly, \textbf{RAG-TR}: RAG Type Resolution).}
    \vspace{-5pt}
    \input{Resources/Tables/rq4-ablation}
    \label{table:rq4-ablation}
\end{table}

\subsubsection{Ablation on I/O Only vs. Full Snapshot Validation}
Against Oxidizer's relaxed criterion (ignoring side-effect/state mutations), I/O-only validation yields 2,560 passes versus full snapshots' 2,451 (Table~\ref{table:rq4-ioonly}); the extra 109 are all \textbf{false positives} masking incorrect side effects (a 4.26\% rate), concentrated in side-effect-heavy projects (\emph{graph}, \emph{validator}).

\begin{table}[t]
    \centering
    \scriptsize
    \setlength{\tabcolsep}{3pt}
    \renewcommand{\arraystretch}{0.95}
    \caption{AMFs passing validation under the full-snapshot oracle vs.\ I/O-only validation; every additional pass is a false positive masking an incorrect side effect.}
    \vspace{-5pt}
    \input{Resources/Tables/rq4-ioonly}
    \label{table:rq4-ioonly}
\end{table}

\mybox{\textnormal{\textbf{Summary.} Ignoring side effects introduces additional false positives (4.26\%), demonstrating that \approach’s full-state checking is necessary to reliably detect incorrect method translations beyond mere input/output equivalence.}}



%% file: Resources/Tables/rq4-ablation.tex
\begin{tabular}{l cccc}
\toprule
\textbf{Subject} & \textbf{\approach} & \textbf{w/o ES} & \textbf{w/o OTF} & \textbf{w/o RAG-TR} \\
\midrule
cli        & 77.29 & 73.26 & 70.33 & 76.92 \\
codec      & 51.76 & 41.76 & 45.00 & 47.65 \\
csv        & 60.85 & 58.30 & 62.13 & 48.09 \\
exec       & 68.95 & 44.35 & 52.02 & 58.06 \\
fast-pfor  & 35.96 & 13.50 & 33.29 & 32.75 \\
fileupload & 73.44 & 29.17 & 41.67 & 66.15 \\
graph      & 43.40 & 27.92 & 39.43 & 38.68 \\
jansi      & 48.41 & 25.67 & 42.79 & 43.28 \\
pool       & 53.37 & 29.62 & 40.62 & 43.70 \\
validator  & 57.59 & 43.34 & 49.23 & 59.44 \\
\midrule
\textbf{Total} & \textbf{52.79} & \textbf{34.96} & \textbf{44.82} & \textbf{47.96} \\
\bottomrule
\end{tabular}

%% file: Resources/Tables/rq4-ioonly.tex
\begin{tabular}{l ccc}
\toprule
\textbf{Subject} & \textbf{Full Snapshot} & \textbf{I/O Only} & \textbf{Masked FPs} \\
\midrule
cli        & 211 & 217 & +6  \\
codec      & 352 & 370 & +18 \\
csv        & 143 & 147 & +4  \\
exec       & 171 & 175 & +4  \\
fast-pfor  & 269 & 275 & +6  \\
fileupload & 141 & 145 & +4  \\
graph      & 230 & 268 & +38 \\
jansi      & 198 & 202 & +4  \\
pool       & 364 & 372 & +8  \\
validator  & 372 & 389 & +17 \\
\midrule
\textbf{Total} & \textbf{2,451} & \textbf{2,560} & \textbf{+109} \\
\bottomrule
\end{tabular}

%% file: Resources/Sections/6-Related-Work.tex
\section{Related Work}
\label{sec:related-work}



Recent work has explored the use of LLMs for code translation~\cite{yan2023codetransocean,yin2024rectifier,jiao2023evaluation,pan2024lost,di2024codefuse,tipirneni2024structcoder,ibrahimzada2026matchfixagent}, showing promising results on synthetic benchmarks but limited effectiveness on real-world software projects. Other studies have similarly employed language models for code migration, with a predominant focus on curated examples~\cite{java2csharp,zhu2022multilingual}. Existing work has also addressed repository-level translation targeting different language pairs. AlphaTrans~\cite{ibrahimzada2025alphatrans} translates Java to Python using open-source LLMs and GraalVM~\cite{graalvm} for isolated validation. Several techniques translate C to Rust~\cite{shiraishi2024smartc2rust,saman2025rustine,shetty2024syzygy}. Oxidizer~\cite{zhang2024scalable} uses fuzzing and FFI-based interoperability, replacing incorrectly translated functions with delegating wrappers rather than isolating per-function validation. It \emph{ignores} instance-state mutations, parameter side effects, and changes to global/static fields, thereby inviting false positives/negatives.
Several methods also integrated transpiler outputs to guide LLM-based translation~\cite{yang2024vert}; agentic pipelines~\cite{ibrahimzada2026recodeagent,ibrahimzada2026matchfixagent,farrukh2026orbit} are now capable of translating repositories. Nitin et al.~\cite{nitin2024spectra} capture natural language specifications from source code to inform translation, while Yang et al.~\cite{yang2024exploring} utilize test cases to support the process.

Compared to these techniques, \approach addresses important limitations in test translation, which many existing techniques rely on to validate the functional equivalence of translations. Our performance is superior to state-of-the-art neurosymbolic and agentic pipelines, notably decreasing false positives and false negatives. 

%% file: Resources/Sections/7-Threats-to-Validity.tex
\section{Threats to Validity}
\label{sec:threats}
\noindent \textbf{External Validity.}
\approach's mechanism is \emph{language-pair agnostic}, relying on \emph{execution-driven semantic serialization and reconstruction via type mapping}. Java$\rightarrow$Python is a particularly challenging static-to-dynamic case, so less-divergent pairs (e.g., Java$\leftrightarrow$Kotlin, Go$\rightarrow$Rust) should generalize more easily, given unit-test and mocking infrastructure is widely available (Python, Rust, Kotlin, C\#, JavaScript).

For sparsely documented libraries, LLM-generated synthetic specifications could replace RAG-retrieved documentation without altering the workflow. Runtime introspection transfers readily to other reflective or dynamic languages (C\#, Kotlin, JavaScript, Ruby), needing extra instrumentation only for lower-level languages (C/C++, Rust). Deserialization into languages like Rust may need structural adaptations (e.g., \texttt{\small{Rc<RefCell<T>>}}), but this is engineering effort analogous to \approach's Java$\rightarrow$Python normalization, not a fundamental barrier.

\noindent \textbf{Internal Validity.}
Risks from incorrect mocks or EvoSuite undercoverage are mitigated by manual inspection of 700 AMFs (confirming a single false positive and far fewer false negatives than GraalVM), an unbiased project pool guarding against project overfitting, and RQ3's cross-LLM (Devstral) results ruling out model overfitting. 

\noindent \textbf{Construct Validity.} \approach relies on well-vetted tools: Tree-sitter~\cite{treesitter}, JaCoCo~\cite{jacoco}, Python coverage~\cite{coverage-lib}, and mock~\cite{mockpy}.

%% file: Resources/Sections/8-Conclusion.tex
\section{Conclusion}
\label{sec:conclusion}

The paper introduced \approach, a framework that integrates context-aware type resolution with mock-based, isolated validation to achieve high-quality, semantically faithful translations from Java to Python. One promising direction for future work is integrating more fine-grained semantic analysis, such as control-flow or data-flow reasoning, to further enhance type inference and translation coherence. Another direction for future work is to make the \approach pipeline PL-agnostic or to enable adaptation to more language pairs with minimal human effort. 

%% file: Resources/Sections/9-Artifacts.tex
\section{Data Availability}
\label{sec:data}

Our artifacts are publicly available at~\cite{website}.